\documentclass[submission,copyright,creativecommons]{eptcs}
\usepackage{geometry}                
\usepackage{breakurl}     
\usepackage{graphicx}
\usepackage{graphbox}
\usepackage{hyperref}
\usepackage{amssymb}
\usepackage{epstopdf}
\usepackage{algorithm}
\usepackage{amsmath}
\usepackage{mathtools}
\usepackage[noend]{algpseudocode}
\usepackage{enumitem}
\usepackage[acronym]{glossaries}
\usepackage[hang,footnotesize,bf]{caption}
\usepackage{cleveref}

\newcommand{\rb}{\mathbf r}
\newcommand{\eb}{x}
\newcommand{\prob}{\mathbb P}
\newcommand{\maxeb}{m}
\newcommand{\teb}{\tilde{\eb}}
\usepackage{listings} 
\usepackage{color}

\definecolor{codegreen}{rgb}{0,0.6,0}
\definecolor{codegray}{rgb}{0.5,0.5,0.5}
\definecolor{codepurple}{rgb}{0.58,0,0.82}
\definecolor{backcolour}{rgb}{0.95,0.95,0.92}

\lstdefinestyle{mystyle}{
    backgroundcolor=\color{backcolour},
    commentstyle=\color{codegreen},
    keywordstyle=\color{magenta},
    numberstyle=\tiny\color{codegray},
    stringstyle=\color{codepurple},
    basicstyle=\ttfamily\footnotesize,
    breakatwhitespace=false,
    breaklines=true,
    captionpos=b,
    keepspaces=true,
    numbers=left,
    numbersep=5pt,
    showspaces=false,
    showstringspaces=false,
    showtabs=false,
    tabsize=2
}

\newacronym{apr}{APR}{annual percentage rate}
\newacronym{bls}{BLS}{Boneh–Lynn–Shacham}
\newacronym{bn}{BN}{Bayesian network}
\newacronym{cbeth}{cbETH}{Coinbase wrapped staked ETH}
\newacronym{cdf}{CDF}{cumulative distribution function}
\newacronym{cl}{CL}{consensus layer}
\newacronym{cpt}{CPT}{conditional probability table}
\newacronym{dvt}{DVT}{distributed validator technology}
\newacronym{eb}{EB}{effective balance}
\newacronym{ef}{EF}{Ethereum Foundation}
\newacronym{eip}{EIP}{Ethereum Improvement Proposal}
\newacronym{el}{EL}{execution layer}
\newacronym{epbs}{ePBS}{enshrined PBS}
\newacronym{ffg}{FFG}{Friendly finality gadget}
\newacronym{fxs}{FXS}{Frax share}
\newacronym{ghost}{GHOST}{Greedy Heaviest-Observed Sub-Tree}
\newacronym{ldo}{LDO}{Lido DAO}
\newacronym{lmd}{LMD}{Latest message driven}
\newacronym{maxeb}{MaxEB}{maximum effective balance}
\newacronym{mev}{MEV}{maximal extractable value}
\newacronym{mm}{MM}{MetaMask}
\newacronym{oobn}{OOBN}{object-oriented Bayesian network}
\newacronym{p2p}{p2p}{peer-to-peer}
\newacronym{pbs}{PBS}{proposer builder separation}
\newacronym{pc}{PC}{personal computer}
\newacronym{pdf}{PDF}{probability density function}
\newacronym{pos}{PoS}{proof of stake}
\newacronym{pr}{PR}{pull request}
\newacronym{rig}{RIG}{Robust Incentives Group}
\newacronym{rpl}{RPL}{Rocket Pool}
\newacronym{ssf}{SSF}{single-slot finality}
\newacronym{steth}{stETH}{Lido staked Ether}
\newacronym{ups}{UPS}{uninterruptable power supply}
\newacronym{vrf}{VRF}{verifiable random function}

\title{Proposer selection in EIP-7251
\footnote{We would like to thank Ben Edgington, Mikhail Kalinin, Barnab\'e Monnot, Mike Neuder and Roberto Saltini for their feedback and discussions. We acknowledge that this research would not have been possible without EF Academic Grant ID: FY23-1030 for which we thank the Ethereum Foundation.}
}

\vspace{16pt}

\author{Sandra Johnson\thanks{Corresponding author}%
	\institute{Consensys Software Inc, Australia}
	\institute{School of Mathematical Sciences, \\ QUT, Australia}
	\email{sandra.johnson@consensys.net}
\and Kerrie Mengersen
	\institute{School of Mathematical Sciences, \\and Centre for Data Science\\ 
	QUT, Australia}
	\email{k.mengersen@qut.edu.au}
\and Patrick O'Callaghan
	\institute{University of Queensland,  Australia}
\and Anders L. Madsen
	\institute{HUGIN EXPERT A/S, Aalborg University \\Denmark} 
	\email{anders@hugin.com}
	}
\date{} 
\begin{document}
\def\titlerunning{Proposer selection in EIP-7251}
\def\authorrunning{S. Johnson, K. Mengersen,  P. O'Callaghan \& A.L. Madsen}
\maketitle
\begin{abstract}
Immediate settlement, or \gls{ssf}, is a long-term goal for Ethereum. The growing active validator set size is placing an increasing computational burden on the network, making \gls{ssf}
 more challenging. \gls{eip}-7251 aims to reduce the number of validators by giving stakers the option to 
  merge existing validators. Key to the success of this proposal therefore is
  whether stakers choose to merge their validators once EIP-7251 is implemented. It is natural
  to assume stakers participate only if they anticipate greater expected
  utility (risk-adjusted returns) as a single large validator. In this paper,
  we focus on one of the duties that a validator performs, viz. being the proposer for the next block. This duty can be quite lucrative, but happens infrequently.  
  Based on previous analysis, we may assume that \gls{eip}-7251 implies no change to the
  security of the protocol. We confirm that the probability of a validator
  being selected as block proposer is equivalent under each consolidation regime. This result
  ensures that the decision of one staker to merge has no impact on the opportunity
  of another to propose the next block, in turn ensuring there is no major systemic change to the
  economics of the protocol with respect to proposer selection.
 \end{abstract}

\section{Introduction}
\label{EIP-7251}
EIP-7251 is a proposal to increase the \gls{maxeb} to 2,048 ETH, which ``allows validators to have larger \glspl{eb}, while maintaining the 32 ETH lower bound'' \cite{Neuder2023e}. 
A key strategy of the proposal is ``validator set contraction'', with the
ultimate aim of advancing, or ideally enabling, the feasibility of \gls{ssf} \cite{buterin2023}
and  enshrined \gls{pbs}, while also reducing ``strain on the \gls{p2p} layer''
\cite{Neuder2023a, Neuder2023e}.

With EIP-7251, validators can choose to auto-compound their stake, instead of
having any rewards above a 32 ETH effective balance automatically withdrawn
into their account.

Therefore if single stake validators change their withdrawal credentials to signify that it is a consolidated auto-compounding validator, every 1 ETH increment of their \gls{eb} from rewards earned will make them eligible for increased rewards based on the larger effective balance, instead of waiting until another 32 ETH of rewards has been accumulated in order to spin up an another validator. Analysis \cite{Edgington2023} has shown a current median of 1.31 ETH/year, a mean of 1.33 ETH with a standard deviation of 0.10 ETH. These calculations were based on an active validator set of 500,000 validators. Other calculations set the estimate to 11 years for an average validator
    to earn 32 ETH
    \cite{Neuder2023a}. The active validator set is currently (15 April 2024) much larger,  being reported as 984,018 by BeaconScan \cite{etherscan} with 21,012 validators in the pending queue. By design, validator rewards decline with an increase in active validator set size.

Increasing the \gls{maxeb} by a factor of 64 from 32 to 2,048 ETH
will result in a less homogeneous validator set with respect to individual
validator weights. The consolidated validators could theoretically have an effective balance ranging from 16 ETH to 2,048 ETH. In the reality, however, the lowest effective balance is currently 28 ETH (15 April 2024) with the overwhelming majority of validators having effective balances of 32 ETH \cite{etherscan}, which is both the minimum required to activate a validator on the beaconchain, and the \gls{maxeb}. If a validator is penalised for not performing their validator duties, they will automatically be exited by the protocol once their effective balance falls below 16 ETH. They are able to initiate top-ups of their validator stake to bring the effective balance back up to 32 ETH.

Currently there is a sweep of the entire validator set to process withdrawals
of all effective balances that are above 32 ETH, the current maximum for
effective balances. EIP-7251 introduces a new \gls{bls} prefix, \texttt{0x02}, so that the
automatic withdrawal for these validators only happens when their effective
balance exceeds a \gls{maxeb} of 2,048 ETH. 

However, for this \gls{eip} to be acceptable to large stakers, they may want to nominate
the amount to partially withdraw ETH from a validator \cite{lidofaq}. For example, perhaps a staker wants to
maintain a consolidated validator that consists of five 32 ETH validators, and
they may want to withdraw rewards in excess of 160 ETH or another arbitrary value, and not the default
value of 2,048 ETH. This potential requirement was recognised and mentioned by Asgaonkar in his
initial proposal \cite{Asgaonkar2023}. \gls{eip}-7002 has been proposed to trigger partial withdrawals from the execution layer \cite{eip-7002} which would satisfy this requirement. Both \gls{eip}-7002 and \gls{eip}-7251 are considered for inclusion in the next hard fork of Ethereum, \href{https://ethereum-magicians.org/t/pectra-network-upgrade-meta-thread/16809}{Electra/Prague(Pectra)}.

In addition to small solo stakers \cite{koliopoulos2023}, compounding stake 
should also be attractive to larger stakers who regularly earn far greater
rewards, and  instead of waiting for a newly created validator to be activated
in the often long  activation queue, that stake will immediately earn additional
rewards. There would also be less overhead when running fewer validators. On the  other hand consolidation and compounding of validators run a
risk of increased slashing penalties.

Consolidation of large stakers would have the most impact on validator set contraction. Consequently the
appetite for consolidation of stake needs to be ratified by stakers, especially the larger stakers.
A security analysis has already been undertaken by the \gls{ef}
\cite{damato2023} which confirms that there is no increased security risk as a result of this \gls{eip}.

The expectation is that the proposer selection mechanism would not need to be altered in EIP-7251 because the probability of proposer selection is influenced by a validator's effective
balance, meaning that validators with a higher stake carry more weight than
those with an effective balance of 32 ETH \cite{Neuder2023a}. However, there is likely to be a slight increase in time to select, since more validators are expected to fail the proposer selection test, therefore requiring more iterations. This is discussed in more detail in the Method section on page \pageref{sec:method}.

Our main research question is: allowing for varied consolidation strategies by large stakers, can we confirm that apart from an expected slight increase in processing time, there is no adverse effect on proposer selection for stakers and validators? 

The motivation  for this research question is three-fold:
\begin{itemize}[noitemsep]
\item ensuring that solo stakers are not inadvertently adversely affected by the introduction of this \gls{eip}.
\item understanding how proposer selection is impacted by the consolidation strategy of larger stakers.
\item estimating the number of iterations required to select the next block proposer.
\end{itemize}
 
We discuss these three aspects of our research question in the \textit{Proposer Selection} section (page \pageref{sec:method}) where we describe the process of selecting the proposer for the next block, followed by the \textit{Bayesian network model} section (page \pageref{sec:results}) where we develop the model and run several scenarios through the model, and lastly we discuss the findings and draw conclusions in the \textit{Discussion and Conclusions} section (page \pageref{sec:discussion}).

\section{Proposer Selection}
\label{sec:method} 

\subsection{Time to select the next proposer}
As noted in the ethresear.ch post \cite{Neuder2023a}, the expectation is that there may be a slight increase in the time it takes to determine the proposer of the next block, due to more iterations of the logic being required due to the lower probabilities of validators with a smaller stake to pass the proposer eligibility test. We confirm this intuition in the graph below where we use a negative binomial to estimate the number of failed proposer eligibility checks before a solo validator passes the check.

\begin{figure}[htbp]
\begin{center}
\includegraphics[width=\linewidth]{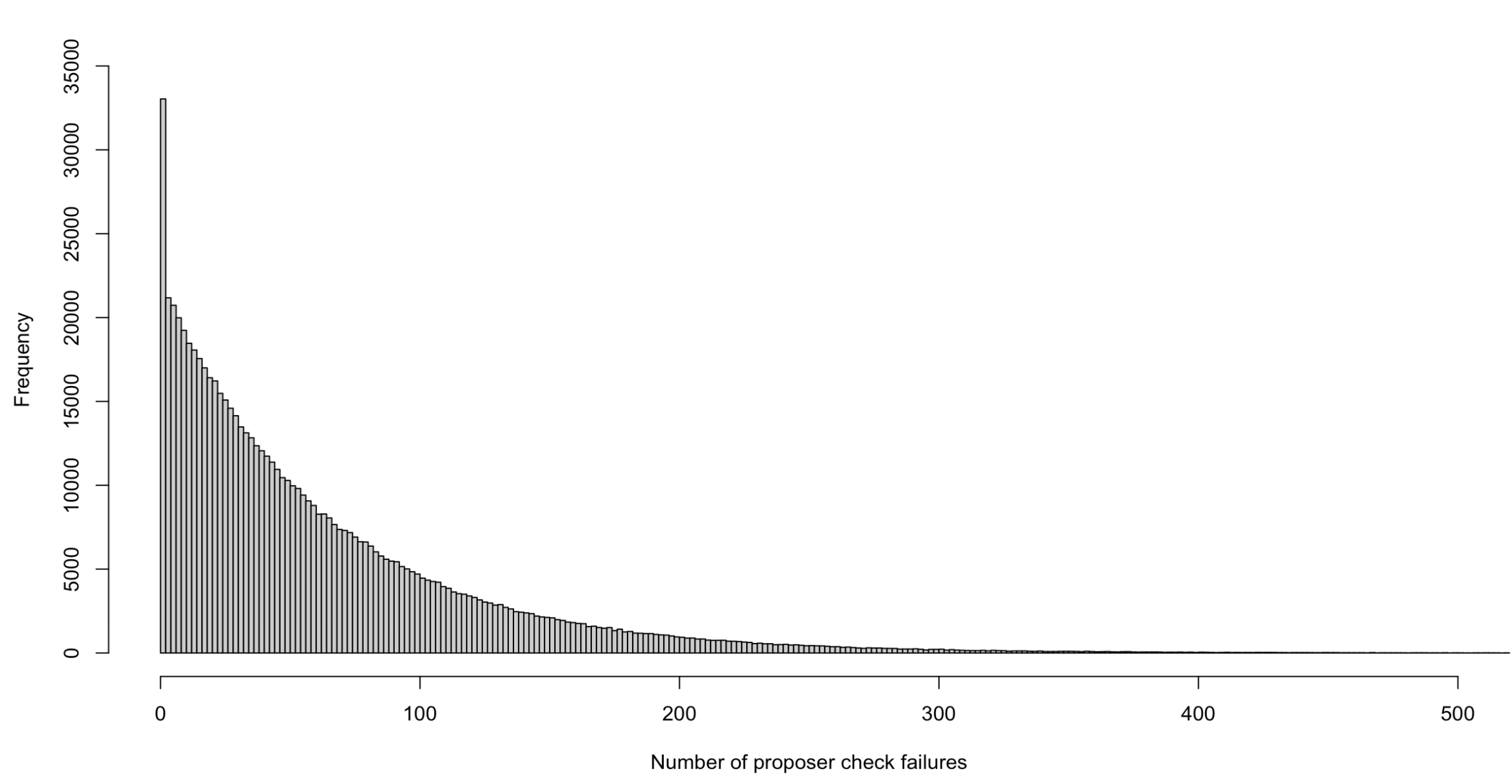}
\caption{Probability distribution of the number of failures of the proposer check before the first successful proposer check}
\label{fig:negbin}
\end{center}
\end{figure}

We only included proposers with 32 ETH effective balances. Therefore the probability of passing the proposer check was 0.016. The graph in \cref{fig:negbin} visualises the probability distribution of the number of failures of the proposer check before the first successful check.

The $median$ value for the number of failures is 43; i.e., we can expect that half of the time more than 43 iterations will be required and half the time fewer than 43 iterations.

Apart from the median, it is also interesting to quantify other probabilities, such as:
\begin{itemize}[noitemsep]
\item Probability of fewer than 100 iterations 
$=$ 0.7962
\item Probability of more than 100 iterations
$=$ 1 - 0.7962 $=$ 0.2038
\item Probability of more than 200 iterations
$=$ 0.0422
\item Probability of more than 300 iterations
$=$ 0.0087
\item Probability of more than 400 iterations
$=$ 0.0018
\end{itemize}

When we iterate through the shuffled validator indices, we observe that before a ``successful'' candidate is reached, all the candidates ahead of the eventual proposer in the shuffled index had to have been rejected.
However, given the large active validator set, the probability calculations based on a large finite sample assumption hold and do not materially change the calculations, and are valid in this case.

These additional iterations should have no noticeable effect on processing time.

\subsection{Proposer selection process}
Proposer selection involves the following steps:
\begin{itemize}[noitemsep]
  \item shuffle the validator set and select the first validator index
  \item generate a random value for the candidate 
  \item check whether the candidate validator passes the \emph{proposer eligibility} check
\end{itemize}

The \emph{swap-or-not-shuffle} technique is used to shuffle the validator
indices in preparation for the selection of the next candidate for a block proposer \cite{hoang2014}. The validator shuffling is done in \href{https://github.com/ethereum/consensus-specs/blob/9c35b7384e78da643f51f9936c578da7d04db698/specs/phase0/beacon-chain.md#compute_shuffled_index}{\texttt{compute\_shuffled\_index}}.

The computation to determine the proposer for the next block is the following (\href{https://github.com/ethereum/consensus-specs/blob/9c35b7384e78da643f51f9936c578da7d04db698/specs/phase0/beacon-chain.md#compute_proposer_index}{\texttt{compute\_proposer\_index}}):

\begin{lstlisting}[language=Python]
def compute_proposer_index(state: BeaconState, indices: Sequence[ValidatorIndex], seed: Bytes32) -> ValidatorIndex:
    """
    Return from 'indices' a random index sampled by effective balance.
    """
    assert len(indices) > 0
    MAX_RANDOM_BYTE = 2**8 - 1
    i = uint64(0)
    total = uint64(len(indices))
    while True:
       candidate_index = indices[compute_shuffled_index(i % total, total, seed)]
       random_byte = hash(seed + uint_to_bytes(uint64(i // 32)))[i % 32]
       effective_balance = state.validators[candidate_index].effective_balance
       if effective_balance * MAX_RANDOM_BYTE >= MAX_EFFECTIVE_BALANCE * random_byte:
           return candidate_index
       i += 1
\end{lstlisting}

Therefore, we iterate through the shuffled indices, starting with the first
entry and then check whether it passes the selection criteria. If it doesn’t,
then the next validator index in the array goes through the same check.

As we can see from the code, the validator’s \gls{eb} is
multiplied by 255 (i.e.  $\texttt{MAX\_RANDOM\_BYTE} = 2^8 - 1 = 255$) and then
compared to the product of the generated random\_byte, $\rb: \Omega \rightarrow
[255]$ and the $\texttt{MAX\_EFFECTIVE\_BALANCE} = 2,048
\textsc{eth}$.\footnote{Here we are adopting the convention that $\Omega$ is
  the implicit state space and $[n]$ is shorthand for the set of integers $0,
\dots 255$.}
  
\begin{figure}[htbp]
\begin{center}
\includegraphics[width=0.95\linewidth]{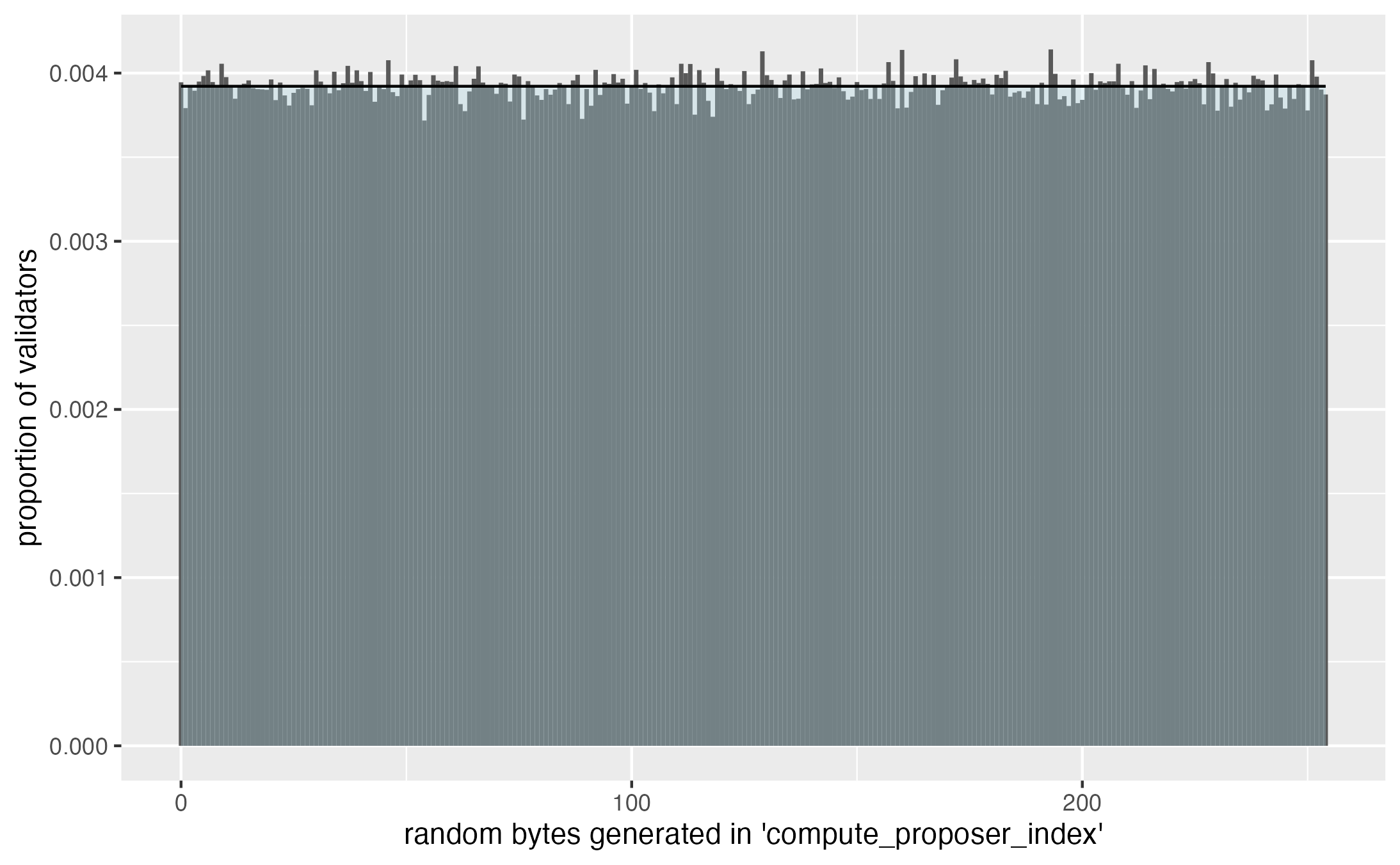}   
\caption{Distribution of 716,800 random bytes generated from the spec}
\label{fig-rb}
\end{center}
\end{figure}

\Cref{fig-rb} is an exposition of random byte values generated for 716,800
validators from the \emph{random\_byte} assignment statement below.
Superimposed on the histogram of these random byte integers is a uniform
distribution. As expected, the random bytes appear to visually resemble values
drawn from a uniform distribution: $\rb \sim U(0,255)$.

\begin{lstlisting}
random_byte = hash(seed + uint_to_bytes(uint64(i // 32)))[i \% 32].
\end{lstlisting}

The probability of a validator being the proposer if their index was selected
from the list is calculated as follows. Let $C_i$ be the event ``candidate $i$
passed proposer check''. Then
\begin{align} \label{eq-candidate}
  C_i = \{\omega : 255 \cdot \eb_i \geq \rb(\omega) \cdot \maxeb \}
\end{align}
$\eb_i = candidate \texttt{ } i \texttt{ } \gls{eb}$ and $\maxeb =
\texttt{MAX\_EFFECTIVE\_BALANCE}$.
Observe that the conditional probability $\prob(C \mid \eb_i = \maxeb) = 1$. In
other words, if candidate $i$'s \gls{eb} is equal to the \gls{maxeb}, then $i$ becomes the proposer with probability 1. Observe that the
above holds, regardless of the value of $\maxeb$.

In the case where $\eb_i < \maxeb$, the probability of passing the proposer
eligibility test will vary depending on the extent of validator consolidation.
Let $\tilde \rb = \rb / 255$ be the normalized random byte function and, for
each $\eb_i \in [32, 2048]$, let $\teb_i \in [1, 64]$, 
$\teb_i = \eb_i / 32$ denote the normalized measure of $i$'s
consolidated effective balance, so that $\tilde \eb_i = 1$ captures the
status quo and $\teb_i = 64$ the maximally consolidated validator. 
Via \cref{eq-candidate},
\[C_i = \{\omega : \rb(\omega) / 255 \leq \eb_i / 2048\}
= \{\omega : \tilde \rb(\omega) \leq \teb_i / 64\},\]
so that, for every $\eb_i \in [32, 2048]$, 
$\prob(C_i \mid \eb_i) = \teb_i /64$,

\begin{table}[htp]
\begin{center}
\caption{Probability candidate $i$ passes the proposer check conditional on
effective balance}
\renewcommand{\arraystretch}{1.3}
\begin{tabular}{c|c|c|c|c|c|c}
  \hline
  $\eb_i$
  & 32 & 64 & 128 & 256 & 512 & 1024\\
  \hline
  $\prob(C_i \mid \eb_i)$&
  0.015625 & 0.03125 & 0.0625 & 0.125 & 0.25 & 0.5\\
  \hline
\end{tabular}
\end{center}
\end{table}
\begin{figure}[htbp]
\begin{center}
\includegraphics[width=0.8\linewidth]{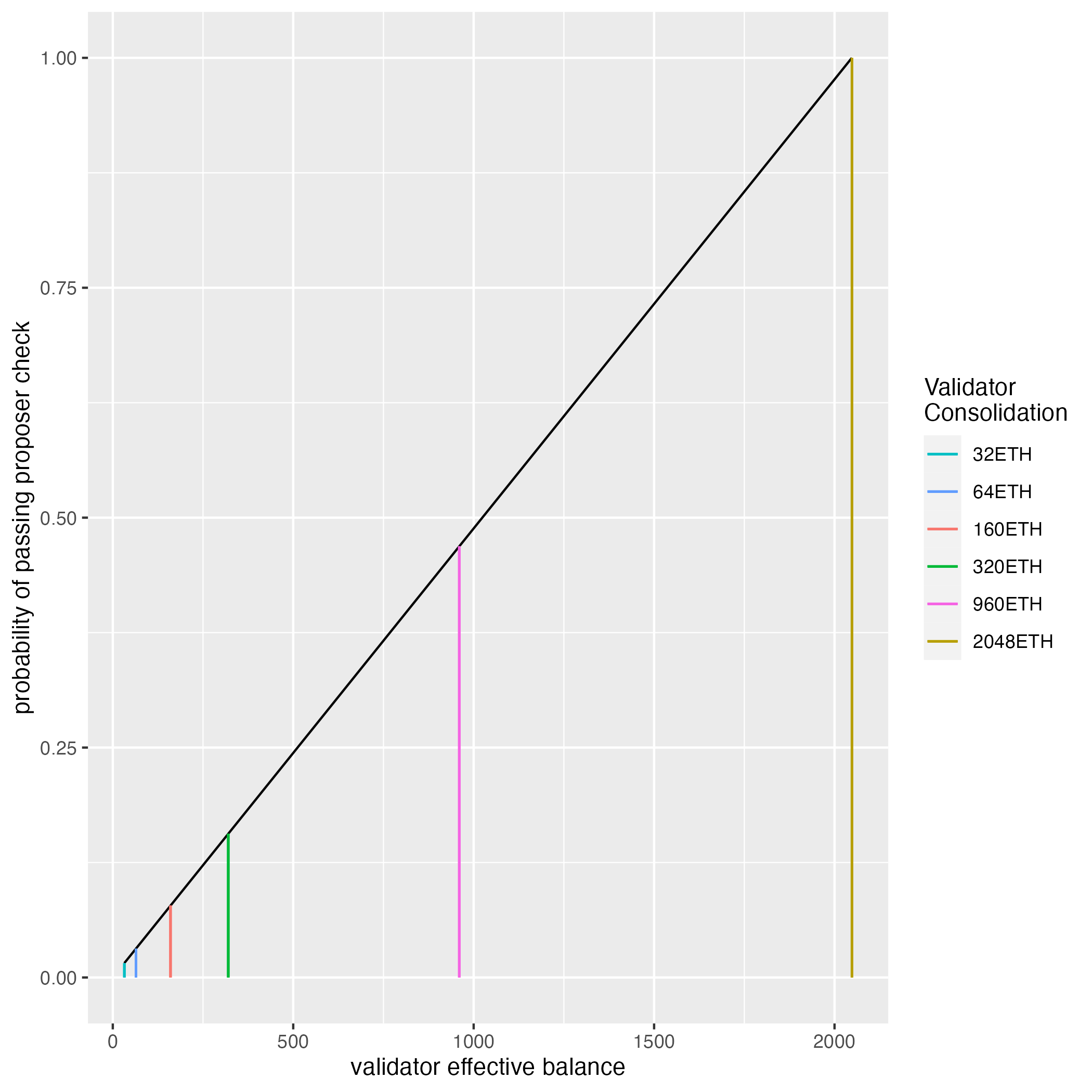}
\caption{Probability of passing the proposer eligibility check for a candidate validator with an EB ranging from 32 to 2,048 ETH}
\label{fig:proposerEB}
\end{center}
\end{figure}
\Cref{fig:proposerEB} illustrates the increased probability of passing the proposer eligibility test as
the validator's effective balance increases.

Assuming a validator set size of 716,800, then the probability of any
validator, regardless of their effective balance, being first in the list of
shuffled validator indices (i.e. the candidate to be assessed as the next block
proposer) is: \\
$\prob(candidate) = \frac{1}{(Active \texttt{ } validator \texttt{ } set \texttt{ }
size)} = \frac{1}{ 716,800} = 1.395 \cdot 10^{-6}$.

From \cref{eq-candidate} we see that currently (i.e. prior to EIP-7251), providing a validator maintains its
effective balance at 32 ETH, once its index is selected as the next candidate
to be checked, it passes the proposer selection test with certainty (i.e.
probability of $1$, or $100\%$).

We calculate the probability that a validator is selected and passes the check for proposer eligibility, as the product of the probability of being the candidate index  and the probability of passing the proposer check, since these two events are independent, i.e. 

$\prob(A \cap B) = \prob(A \mid B)\prob(B) = \prob(A)P(B)$ \\
Putting it another way:
Given \gls{eb}=32 ETH, then the probability of being selected as the $1^{st}$
candidate and becoming the next proposer is calculated as follows: \\
$\prob(candidate$ $\&$ $proposer$ $check$ $passed)$  $= \prob(candidate) \cdot
\prob(proposer$ $check$ $passed \mid candidate) = $ 
$\prob(candidate) \cdot \prob(proposer$
$check$ $passed) = \left(\frac{1}{716800}\right) \cdot 1 \approx 1.395 \cdot 10^{-6}$\\

\noindent
If MaxEB = 2,048 ETH, this changes to: \\
$\prob(candidate$ $\&$ $proposer$ $check$ $passed) =$
$\left(\frac{1}{716800}\right) \cdot \left(\frac{32}{2048}\right)$
$\approx 2.18\cdot 10^{-8}$ \\

\noindent
Consequently, the selected candidate at any round will be less likely to pass the proposer check, especially if they have an \gls{eb} of 32 ETH or lower. Therefore, 
the validator's index which was selected as the next block proposer
may quite likely have been selected at a later round, e.g. $2^{nd}, 3^{rd}, ..., n^{th}$.

We calculate the probability that a 32 ETH validator, $v$, is
selected at round $i$ given that all the previous rounds $(1,2, ... ,(i-1))$
were unsuccessful, i.e. we sum over all the possible rounds that this validator
may have been selected.  \\

\noindent
$\prob(validator \texttt{ } v \texttt{ } selected \texttt{ } as
\texttt{ } next \texttt{ } proposer) = \sum_{i=1}^{716,800} p_i \cdot  \left(
\prod_{j=1}^i (1-p_{j-1}) \right), \texttt{ } where$ \\
$p_i = \prob(round \texttt{ }i \texttt{ } proposer) = \prob(round \texttt{ } i \texttt{
} candidate) \cdot  \prob(passing \texttt{ } proposer \texttt{ } eligibility)$, $and$\\
$p_0 = 0$. This probability is also $1.395 \cdot  10^{-6}$.

In summary, with \gls{eip}-7251, a solo staker is selected as the
next proposer with a probability of $2.18\cdot 10^{-8}$, from the current $1.395 \cdot 
10^{-6}$ probability if they are the first candidate in the shuffled index.
However, their probability over all the possible outcomes will be the same as
is currently the case, assuming all the other validators have the same \gls{eb}.

The probability calculations above assume a validator set with no
consolidation, i.e. the validator set size remained unchanged after the
introduction of \gls{eip}-7251. In practice this is highly unlikely. 

The landscape would be far more interesting if we considered a less homogenous distribution of validator sizes.
Therefore, let us consider an example scenario:\\

\noindent
\textbf{Example Scenario}: \emph{Active validator set has validators with
    varying EBs (i.e. stakers may choose different strategies for validator
    consolidation). To the best of our knowledge, the analyses to date have
    mainly been for a homogeneous validator set, i.e. all validators have the
    same effective balance, either unconsolidated (32 ETH) or fully
    consolidated (2,048 ETH). This includes the analysis we conducted above.}\\

Using this example we build an \gls{oobn} to represent the scenario and glean some insights into proposer selection \cite{KjaerulffMadsen2013}.

\section{Bayesian network model}
\label{sec:results}
It is possible to stake in Ethereum in a number of ways \cite{breslina, efstaking, efpools, efsaas, efsolo}. 

We group the stakers into five categories:
\begin{enumerate}[noitemsep]
\item Small-scale solo stakers (32 - a few hundred ETH)
\item Large-scale individual solo stakers (1000+ ETH)
\item Large-scale institutional solo stakers (ie. companies staking their own ETH)
\item Centralised staking pools
\item Semi-decentralised staking pools (e.g. Rocketpool \cite{RocketPool2023}, Lido \cite{lidodoc}... )
\end{enumerate}

\begin{figure}[htbp]
\begin{center}
\includegraphics[width=0.95\linewidth]{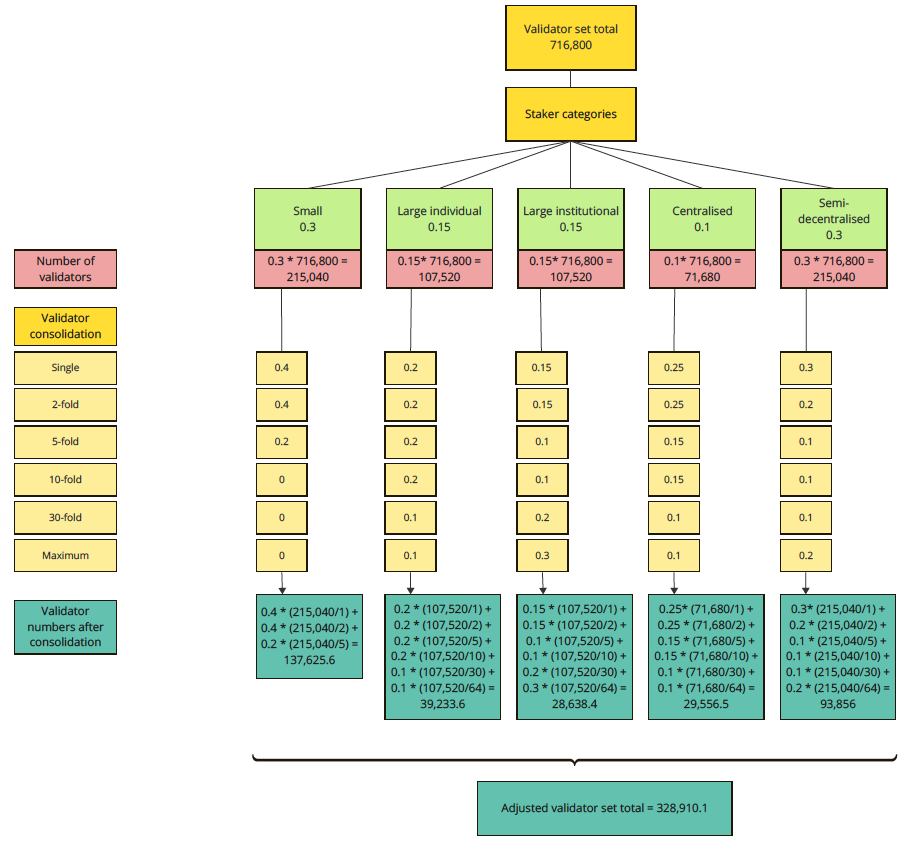}
\caption\emph{Visual representation of the example scenario for validator
consolidation}
\label{fig:consolidation}
\end{center}
\end{figure}

In our hypothetical example in \cref{fig:consolidation} we assume that we have an active validator set with 716,800
validators prior to \gls{eip}-7251. The stakers in this validator set may be categorised as mentioned above, with each category having a distinct consolidation strategy. 

In \cref{fig:consolidation}  we see that the proportion of small stakers is 30\%, large individual stakers 15\%, large institutional 15\%, centralised staking pools 10\% and semi-decentralised staking pools 30\% (light green rectangles). The number of validators run by each staker category is shown as light red rectangles below each category.

The hypothetical consolidation strategy for each of these categories is shown in \cref{fig:consolidation} and depicts the proportion of each consolidated validator type (yellow rectangles) and the resulting number of validators they have after implementing the consolidation strategy  (cyan rectangles), with the total adjusted validator set size following all the consolidations in the bottom rectangle i.e., 328,910.

We put a potential distribution over the various validator consolidations in our scenario for illustrative purposes. The distributions can be adjusted as required.

The \emph{Validator numbers after consolidation} row (green) shows the reduced validator numbers for each category, using the consolidations shown in each column (yellow).

Based on the chosen configuration, the total validator set size reduces to 329,810 (last value in the diagram).

We built a \gls{bn} to illustrate the dependencies between the different factors in the BN model for the example scenario described above (Figures 4 and 5) \cite{Marcot2007, KjaerulffMadsen2013}.

A \gls{bn} is an acyclic directed graph of nodes and edges. The nodes represent the key factors of the system or problem being modelled, and the edges between the nodes indicate dependencies. Uncertainty and the strength of the dependencies between connected nodes are explicitly captured in the node probability tables that are attached to each node in the model. An object-oriented version of a BN (OOBN) may be used to make BNs less cluttered and more readable, by grouping related nodes and processes in OOBN submodels \cite{KollerPfeffer1997, Johnson2013}.

In \glspl{bn} and \glspl{oobn} there are various node types, but the ones relevant to the \gls{oobn} model we developed are: private nodes (not visible to any other models), e.g. \textit{Staker Categories}, instance nodes (an instance of another \gls{oobn} model), e.g. \textit{Probability calculations} and interface nodes (two types: output nodes, e.g. \textit{Validator selected as candidate for proposer duty} and input nodes, e.g. \textit{Consolidated validator types}). The interface nodes define the way in which other models can interact with the \gls{oobn} \cite{Johnson2012a}.

\begin{figure}[htbp]
\begin{center}
\includegraphics[width=0.29\linewidth]{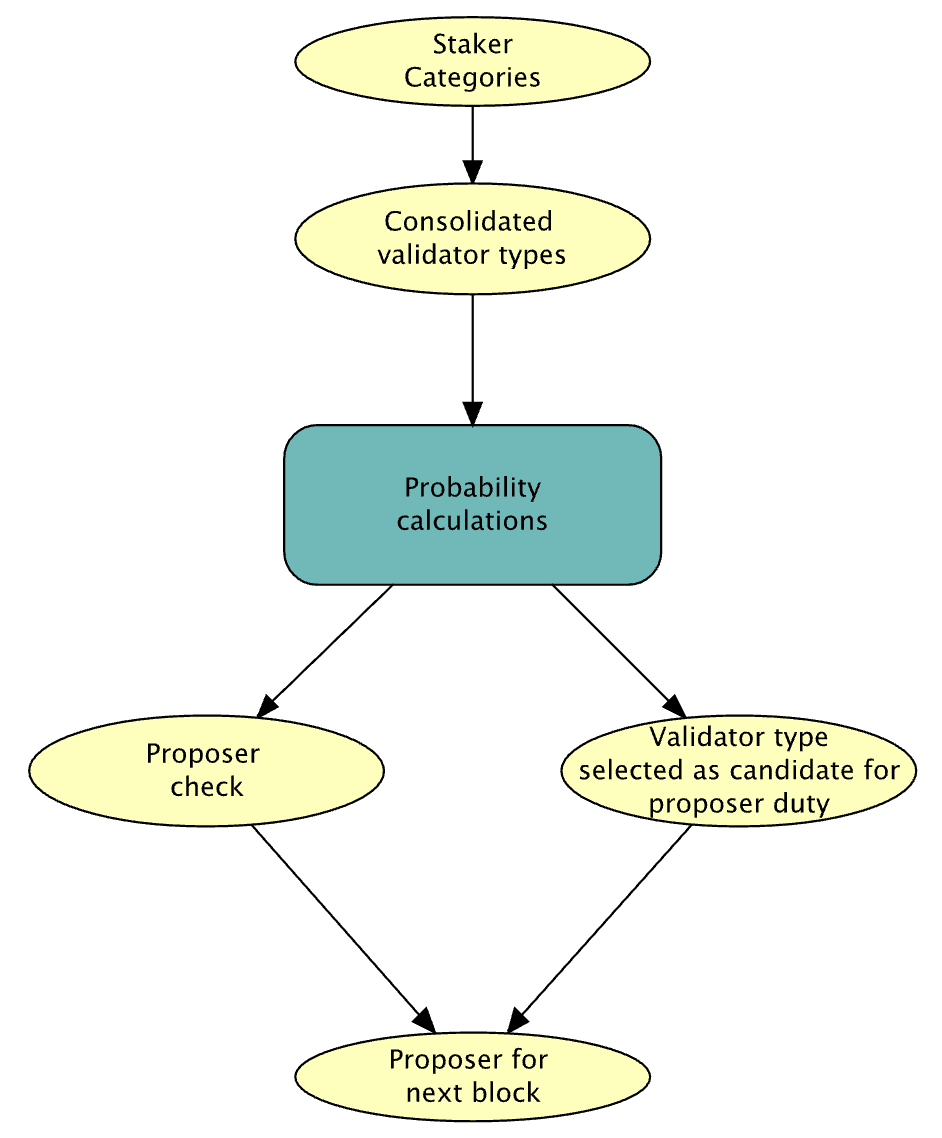}
\includegraphics[width=0.69\linewidth]{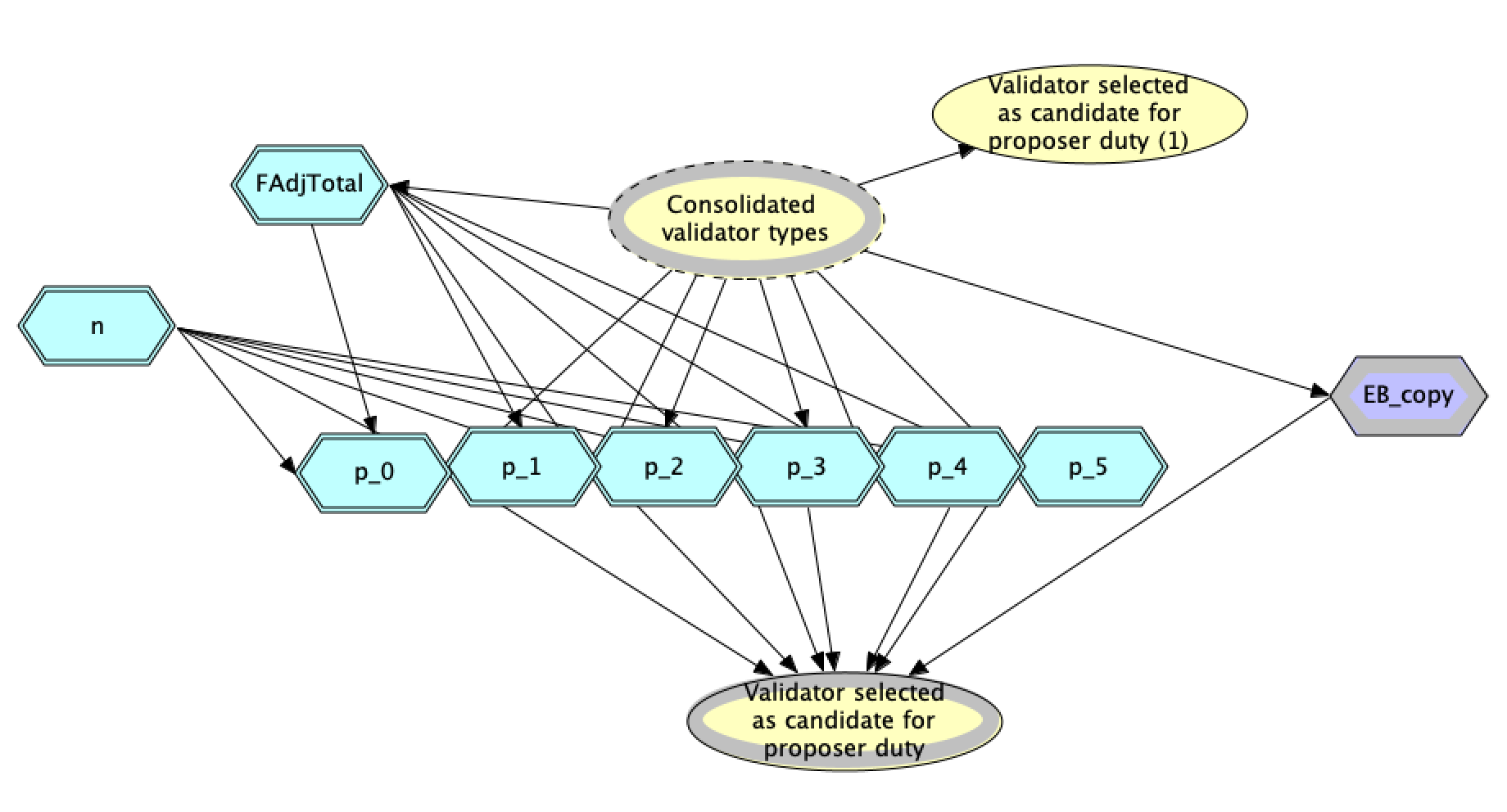}
(a) \hspace{200pt} (b)
\caption{(a) OOBN for proposer selection (b) OOBN subnet for probability calculations}
\label{fig:oobn}
\end{center}
\end{figure}

The probabilities shown in \cref{fig:proposerEB} are used in the node probability table for BN node \textit{Proposer check}, which captures the probability that a validator will pass the proposer eligibility check for the various consolidated validator sizes.

\Cref{fig:oobn} (a) is the high level \gls{oobn} 
\footnote{An \href{https://demo.hugin.com/example/ProposerSelection}{online version} of the proposer selection \gls{oobn} model is freely available. This research has been shared with the Ethereum community in a \href{https://ethresear.ch/t/proposer-selection-with-increased-maxeb-eip-7251/18144}{blog} at ethresear.ch }
\begin{table}[htp]
\begin{center}
\caption{Probability that validator type is selected as a candidate proposer}
\renewcommand{\arraystretch}{0.9}
\begin{tabular}{c|c|c|c|c}
  \hline
   \textit{Type of}           & \textit{Proportion}     & \textit{Number of}    & $\prob(validator \texttt{ } type$ & $\prob(selected$ \\
   \textit{validator}         & \textit{of validators}   & \textit{validators in} & $selected \texttt{ } as$              & $validator$ \\
   \textit{consolidation}  & \textit{consolidating} & \textit{consolidated} & $candidate)$                            & $type \texttt{ } is \texttt{ } proposer)$ \\
                                     &                                   & \textit{validator set} &                                                   & $proposer)$ \\
  \hline
 Single & 28.75\% & $\frac{0.2875*716,800}{1}$ & $\frac{206,080}{329,810} = 0.6248$ & $0.6248 * 0.016$ \\
            &               & $=206,080$                          &                                                            & $=0.0100$ \\
  \hline
  Partial (2-fold)     & 25.75\% & $\frac{0.2575*716,800}{2}$ & $\frac{92,288}{329,810} = 0.2798$ & $0.27988 * 0.031$ \\
            &               & $= 92,288$                                            &                                                         & $=0.0087$ \\
  \hline
  Partial (5-fold)     & 15.00\% & $\frac{0.1500*716,800}{5}$ & $\frac{21,504}{329,810} = 0.0652$ & $0.0652 * 0.078$ \\
            &               & $=21,504$                          &                                                            & $=0.0051$ \\
  \hline
  Partial (10-fold)   & 9.00\% & $\frac{0.0900*716,800}{10}$ & $\frac{6,451}{329,810} = 0.0196$ & $0.0196 * 0.156$ \\
            &               & $= 6,451$                                            &                                                         & $=0.0031$ \\
 \hline
  Partial (30-fold)   & 8.50\% & $\frac{0.0850*716,800}{30}$ & $\frac{2,031}{329,810} = 0.0062$ & $0.0062 * 0.469$ \\
            &               & $= 2,031$                                            &                                                         & $=0.0029$ \\
  \hline
  Partial (64-fold)   & 13.00\% & $\frac{0.1300*716,800}{64}$ & $\frac{1,456}{329,810} = 0.0044$ & $0.0062 * 0.469$ \\
            &               & $= 1,456$                                            &                                                         & $=0.0044$ \\
  \hline
   &  &   &   &  \\
 \textbf{TOTAL}   & \textbf{100.00\%} & \textbf{329,100} & \textbf{1.0000} &  \\
  \hline

\end{tabular}
\end{center}
\end{table}

\subsection{Running the \gls{oobn} model}
\begin{figure}[htbp]
\begin{center}
\includegraphics[width=0.3\linewidth]{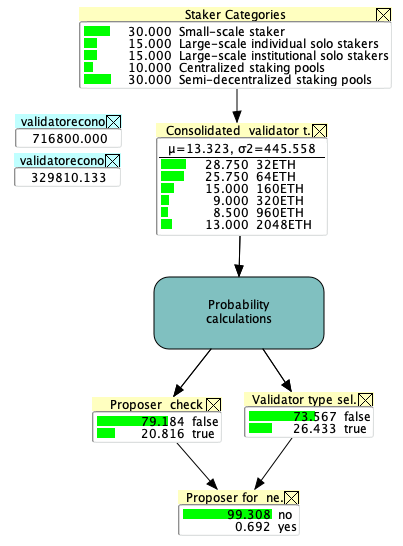}
\includegraphics[width=0.3\linewidth]{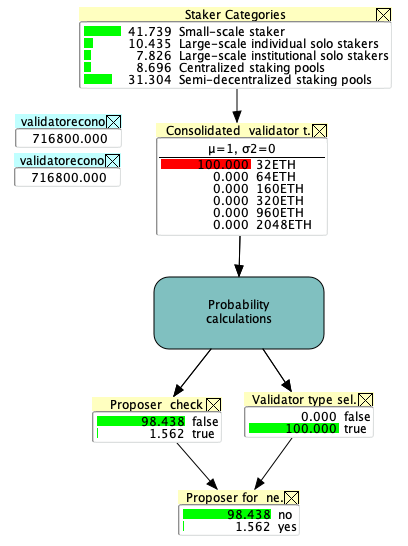}
\includegraphics[width=0.3\linewidth]{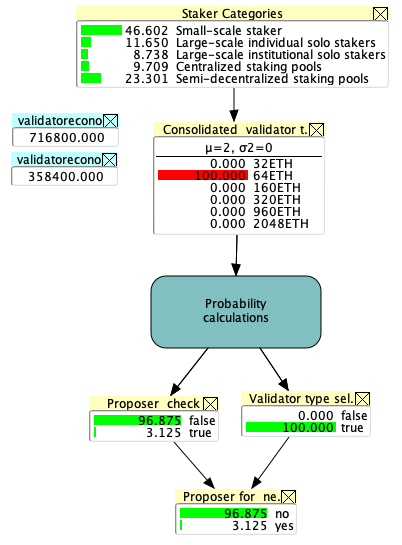} \\
\hspace{20pt}(a) \hspace{110pt} (b) \hspace{110pt} (c) \\
\includegraphics[width=0.3\linewidth]{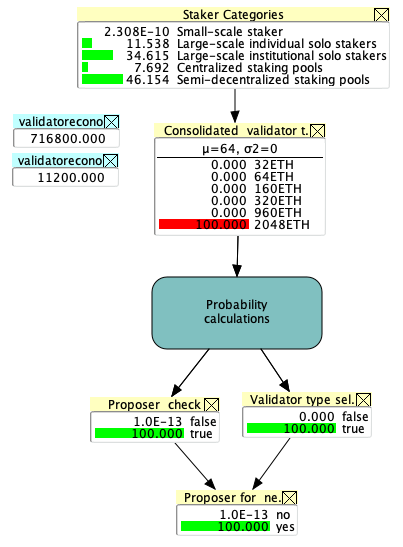}
\includegraphics[width=0.3\linewidth]{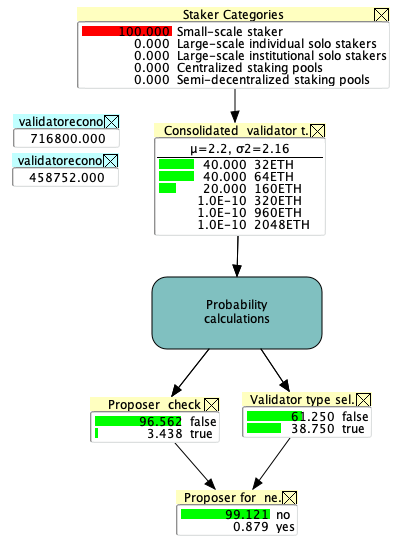}
\includegraphics[width=0.3\linewidth]{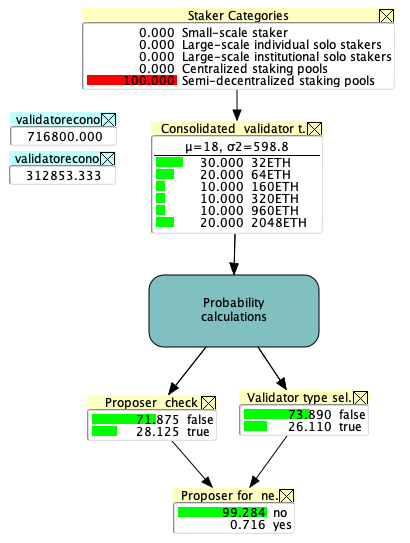} \\
\hspace{20pt}(d) \hspace{110pt} (e) \hspace{110pt} (f) \\
\caption{Running various scenarios through the proposer selection OOBN model}
\label{fig:runoobn}
\end{center}
\end{figure}

\subsubsection*{Scenario 1: No evidence entered into the model (\Cref{fig:runoobn} (a))}
The result of running the \gls{oobn} model is shown in \cref{fig:runoobn} (a). Therefore, assuming the active validator set consists of the staking categories as described along with their respective example consolidation strategies, the validator set will be reduced to 329,810 from 716,800. This adjusted validator set has different proportions of consolidated validators as shown in  \textit{Consolidated validator types}. Across this validator set, the proportion of validators that will pass the proposer test is 20.8\% and the marginal probability, across the various validator types in the validator set, of being selected as the first candidate is 26.4\%, and of being the proposer of the next block is 0.69\%.

The probability of being the next proposer depends on being selected as the next candidate and then passing the test. It appears counter-intuitive that the probability is so much smaller than the individual probabilities in the BN. The reason for this becomes clearer when we look at the conditional probability table of this node, \Cref{tbl:cpt}:

\begin{table}[htp]
\caption{\gls{cpt} for \textit{Proposer for next block}}
\begin{center}
\begin{tabular}{c|c|c|c|c}
  \hline
  Validator type selected as candidate for proposer duty & \multicolumn{2}{|c}{false} & \multicolumn{2}{|c}{true} \\
  \cline{2-5}
  Proposer check    & false & true & false & true \\
  \hline
  $\prob(\textit{Proposer for next block} = no) $    & 1 & 1 & 1 & 0 \\
   $\prob(\textit{Proposer for next block} = yes)$   & 0 & 0 & 0 & 1 \\
  \hline
\end{tabular}
\end{center}
\label{tbl:cpt}
\end{table}

\subsubsection*{Scenario 2: All validators have an effective balance of 32 ETH  (\Cref{fig:runoobn} (b))}
If we enter evidence (shown as a state of a node being red) that the validator set consists entirely of validators with 32 ETH, there is no reduction in the size of the validator set, and the probability of selecting a solo stake validator type is understandably 100\%, with the probability of passing the proposer check being 1.56\%. Hence the probability of a 32 ETH validator type being the proposer of the next block is simply the probability of passing the proposer check, which is 1.56\%.

\subsubsection*{Scenario 3: All validators have an effective balance of 64 ETH (\Cref{fig:runoobn} (c))}
In this scenario we assume that all stakers decided to consolidate their validators into 64 ETH validators. Therefore the validator set size reduces by half to 358,400. As in Scenario 2, the validator type of the first selected candidate will be a validator with 64 ETH, i.e. 100\%. For a validator with an EB of 64 ETH, the probability of passing the proposer check is 3.12\%, meaning that the probability of a validator type of 64 ETH being the next proposer is 3.12\%.

\subsubsection*{Scenario 4: All validators have an effective balance of MaxEB (2,048 ETH) (\Cref{fig:runoobn} (d))}
In this scenario we assume that all stakers decided to consolidate their validators to the maximum allowed, i.e. 2,048 ETH. This is the largest reduction in validator set size, being just 11,200. As in Scenarios 2 and 3, the first selected validator type will definitely be a validator with an \gls{eb} of \gls{maxeb}. For MaxEB the probability of passing the proposer check is 3.12\%, meaning that the probability of a validator type of 64 ETH being the next proposer is 3.12\%.

\subsubsection*{Scenario 5: The validator set consists entirely of small-scale stakers applying the example consolidation strategy (\Cref{fig:runoobn} (e))}
The small staker group is assumed to mainly consist of validators with 32 or 64 ETH, with a small proportion consolidated into 160 ETH validators. Based on this strategy, the validator set reduces to 458,752. Across these small staker validators, the proportion that will pass the proposer test is 3.44\% and the marginal probability, across the three validator types in the validator set, of being selected as the first candidate is
38.75\%, and of being the proposer of the next block is 0.88\%.

\subsubsection*{Scenario 6: The validator set consists entirely of semi-decentralized staking pools applying the example consolidation strategy (\Cref{fig:runoobn} (f))}
In the example consolidation strategy for semi-decentralized staking pools, they are assumed to have a fairly even spread across the various extents of consolidation, with a slight majority of validators being single stake validators (32 ETH). Based on this strategy, the validator set reduces to 312,853. Across this validator set, the proportion of validators that will pass the proposer test is 28.13\% and the marginal probability, across the three validator types in the validator set, of being selected as the first candidate is
26.11\%, and of being the proposer of the next block is 0.72\%.

\section{Discussion and Conclusions}
\label{sec:discussion}
At the time of writing, \gls{eip}-7251 is included in the list of upgrades for the next hardfork of Ethereum -  \href{https://ethereum-magicians.org/t/pectra-network-upgrade-meta-thread/16809}{Electra / Prague (Pectra)}. The `dilemma' facing stakers would be whether they should consolidate and to what extent they should consolidate. Moreover, staking pools would need to adjust their processes to maximise any benefit from this \gls{eip} and importantly put the necessary checks and balances in place to avoid or mitigate any potential risks.

The probability of being selected as the candidate from the shuffled consolidated validator set of size n is the same for each validator, regardless of the extent of consolidation, viz. $\frac{1}{n}$.

So if staker A has 64 single validators and staker B has one consolidated staker, then the probability that a validator from staker A or staker B is the next proposer is calculated as follows: \\

\noindent
$\prob(\textit{staker A is the next proposer}) = \frac{64}{n} * \frac{32}{2048} = \frac{1}{n}$ \\
\noindent
$\prob(\textit{staker A is the next proposer}) = \frac{1}{n} * 1 = \frac{1}{n}$ \\

From this we can deduce that for this scenario, as far as proposer selection is concerned, it is equally likely for a large staker that one of their validators will be the next proposer, regardless of whether they decide to consolidate validators to the full \gls{maxeb}, or leave them all as unconsolidated.

However, there are several other considerations for a staker to take into account when deciding on a consolidation strategy, such as compounding, rewards earned, slashing risk and penalties that vary with effective balance. Proposer selection is just one part of the puzzle.

\nocite{*}
\bibliographystyle{eptcs}
\bibliography{references}

\end{document}